\documentclass[a4paper,fleqn]{cas-sc}

\usepackage[numbers]{natbib}
\usepackage{wrapfig}
\usepackage{gensymb}
\def\tsc#1{\csdef{#1}{\textsc{\lowercase{#1}}\xspace}}
\tsc{WGM}
\tsc{QE}
\tsc{EP}
\tsc{PMS}
\tsc{BEC}
\tsc{DE}

\begin{document}
\let\WriteBookmarks\relax
\def\floatpagepagefraction{1}
\def\textpagefraction{.001}
\shorttitle{Leveraging social media news}
\shortauthors{P. Singh et~al.}

\title [mode = title]{Steady-State Models of STATCOM and UPFC using Flexible Holomorphic Embedding}                      



\author[1]{Pradeep Singh}[type=editor,
                        auid=000,bioid=1,
                        prefix=,
                        role=,
                        orcid=]
\ead{psn121988@gmail.com}


\address[]{Department of Electrical Engineering, Indian Institute of Technology, Delhi,
30332 INDIA}


\author[2]{Nilanjan Senroy}[%
   role=,
   suffix=,
   ]
\ead{nsenroy@ee.iitd.ac.in}







\begin{abstract}
To investigate the effect and ability of FACTS devices using Fast and Flexible Holomorphic Embedding technique (FFHE), it is necessary to develop an embedded system for these devices. Therefore, this paper presents FFHE based embedded system for STATCOM and UPFC. The embedded system is also proposed for their controlling modes. 

The introduced embedded system for STATCOM and UPFC is flexible which allows to take any state as an initial guess instead of fixed state, which leads towards the reduced runtime and decrease the required number of terms, as compared to standard Holomorphic Embedded Load-Flow method (HELM). To demonstrate the effectiveness and practicability, the proposed models of STATCOM and UPFC have been tested for several cases. Further, the developed recursive formulas for power balance equations, devices' physical constraints and their controlling modes are thoroughly investigated and examined. From several tests, it is found that the proposed embedded system requires less execution time and reduce the error at higher rate.
\end{abstract}




\maketitle

\section{Introduction}
FACTS controllers play vital roles in diversified domains of power system operation and control. These devices improve voltage profile, transient stability, voltage stability, available transfer capability \textit{etc.}, thereby improving the overall performance of the power system. But their optimal installation and performance analysis is indispensable to harness maximum benefit-cost ratio. Therefore, their accurate steady-state mathematical models are required to emulate their functionality and performance. These models are integrated with non-linear static load-flow equations of the power systems, which increases the complexity of load-flow problem \cite{zhang2003advanced, zhang2003modelling, wei2004common, jiang2008novel, singh2017amalgam}.

The conventional steady-state models of FACTS controllers are based on numerical iterative techniques \textit{e.g.} Gauss-Seidal (GS), Newton-Raphson (NR), Fast Decoupled (FD) \textit{etc}. These iterative techniques are primarily constrained by the requirement of a good initial guess and they offer slow convergence or even divergence in some cases. Additionally, the solution provided by these methods is indecisive, because sometimes these techniques diverge due to a bad initial guess (although the solution exists) and sometimes converge to a spurious solution (even when the solution does not exists). These limitations were overcome by the Holomorphic Embedding Load-Flow Method (HELM) developed in \cite{trias2012holomorphic}. 

The seminal work in development of HELM framework has been presented in \cite{trias2012holomorphic, rao2016holomorphic, rao2017estimating, wallace2016alternative, subramanian2013pv, trias2016holomorphic, liu2017real}; and advanced applications of HELM in power system analysis can be found in \cite{yao2018voltage, rao2019three, yao2019efficient, wu2019holomorphic, singh2020extended, asl2019holomorphic, freitas2019restarted, li2018numerical}. The main advantage of HELM are deterministic initial guess and assured convergence if the solution exists. However, it has been reported that HELM requires 10 to 20 times more execution time than NR method \cite{sauter2017comparison}. Therefore, Fast and Flexible Holomorphic Embedded (FFHE) method to reduce the execution time has been proposed in \cite{chiang2017novel}. FFHE provides the flexibility to set any arbitrary state as an initial guess. This property circumvents the need to restart the load-flow program when bus-type switching takes place. Therefore, FFHE requires less number of terms to converge if initial guesses are promising. Hence, FFHE is fast and more suitable for load flow analysis than conventional iterative methods.

HELM framework comprising of HE models of thyristor-based FACTS controllers have been proposed in \cite{basiri2017holomorphic}. Unified Power Flow Controller (UPFC) is considered as the superior controller among all FACTS controllers \cite{zhang2003modelling, wei2004common, jiang2008novel}. UPFC is a VSC-based FACTS controller which are more advanced and sophisticated as compared to Thyristor-based FACTS controller because they inject less harmonics, provide independent control, fast dynamic response and higher flexibility. In \cite{singh2019statcom}, a VSC-based FACTS controller, namely Static Synchronous Compensator (STATCOM), has been developed as a variable voltage source using standard HELM. However, the presented model of STATCOM requires larger runtime and also doesn't offers the flexibility to use any state as the starting point. Therefore, the main aim of this work is to develop the FFHE based models of VSC-based FACTS controllers. In this research work, FFHE based models of two VSC-based FACTS controllers, namely Static Synchronous Compensator (STATCOM) and Unified Power Flow controller (UPFC) have been developed along-with their controlling modes.

The remaining part of this paper is organised as follows: Section \ref{PM} deals with the methodology used to develop the FFHE based models of STATCOM and UPFC along-with their controlling modes. In Section \ref{results}, the key findings and numerical results of the introduced embedded system have been discussed. Section \ref{conclusions}, presents the conclusion of this study.

\section{Proposed Holomorphic Embedding Formulations} \label{PM}
This section addresses the procedure adopted to develop a flexible embedded system for STATCOM and UPFC. During development of steady-state model, it is assumed that the system and stated devices are 3-phase balanced and the harmonics generated by them are negligible. Firstly, the embedded system and recursive relationships for the STATCOM and UPFC have been developed and finally, numerical values of all unknown variables along-with their operation bounds are investigated. 

\subsection{STATCOM Modeling}    \label{STATCOM_modeling}
\begin{figure}
\centering
\includegraphics[scale=1.4]{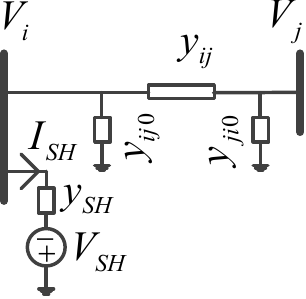}
\caption{Equivalent circuit of a STATCOM}
\label{ST_Equivalent}
\end{figure}

A STATCOM is a shunt connected FACTS device acts as a controllable voltage source. In steady-state, it can inject or absorb reactive power by regulating the injected voltage source magnitude and phase angle. Generally, it consists of a voltage source converter, a capacitor and a coupling transformer. The shunt connected FACTS devices are mainly used to regulate the bus voltage magnitude, although these devices can also be used to control other parameters of the system as discussed in \cite{zhang2004multi}. Figure \ref{ST_Equivalent} shows the topological structure of a STATCOM. 

In Figure \ref{ST_Equivalent}, $y_{SH}$, $I_{SH}$ and $V_{SH}$ are the coupling transformer leakage impedance, complex current and voltage injected by the STATCOM at bus $i$ respectively. The required equations to represent the STATCOM are the power balance equation (PBE) at the connected bus and two equations from device's physical constraint: real power injection criteria and control mode. Equations (\ref{ST_Node}), (\ref{ST_Active_Cons}), and (\ref{ST_Modes}) provide the PBE at $i^{th}$ bus, real power injection criterion and various control modes for STATCOM. Equations (\ref{ST_Mode1}), (\ref{ST_Mode2}), (\ref{ST_Mode3}), (\ref{ST_Mode4}), and (\ref{ST_Mode5}) represent the control constraint equations for the bus voltage magnitude, injected voltage phasor magnitude, injected reactive power, reactive power-flow between buses and equivalent imaginary admittance respectively. 
\begin{equation}
V^{*}_{i}\sum_{k=1}^{N} Y_{ik}V_{k} + V^{*}_{i}(V_{i}+V_{SH})y_{SH}= S^{*}_{i}
\label{ST_Node}
\end{equation}
\begin{equation}
\Re \big [ V_{SH} ( V^{*}_{i}+V^{*}_{SH})y^{*}_{SH} \big ]= 0
\label{ST_Active_Cons}
\end{equation}
\begin{subequations}
\begin{equation}
V_{i}V^{*}_{i}  = |V^{SP}_{i}|^2 
\label{ST_Mode1}
\end{equation}
\begin{equation}
V_{SH}V^{*}_{SH}  = |V^{SP}_{SH}|^2 
\label{ST_Mode2}
\end{equation}
\begin{equation}
\Im \big [ V_{SH}( V^{*}_{i}+V^{*}_{SH})y^{*}_{SH} \big ]  = Q^{SP}_{SH} 
\label{ST_Mode3}
\end{equation}
\begin{equation}
\Im \big [ V_{i}( V^{*}_{i}-V^{*}_{j})y^{*}_{ij} \big ]  = Q^{SP}_{ij} 
\label{ST_Mode4}
\end{equation}
\begin{equation}
\Im \bigg [ \frac{( V_{i}+V_{SH})}{V_{SH}}y_{SH} \bigg ]  = b^{SP}_{eq(SH)} 
\label{ST_Mode5}
\end{equation}
\label{ST_Modes}
\end{subequations}
where, $V_{i}$, $S_{i}$ and $Y_{ik}$ are the voltage phasor of bus $i$, injected complex power at bus $i$ and $(i,k)^{th}$ element of the bus admittance matrix respectively. The quantities with suffix $SP$ are the specified or targeted values. The symbols $N$, $N_{PV}$, and $N_{PQ}$ denotes the set of total, generator and load buses respectively. The operators $\Re(\bullet)$, $\Im(\bullet)$ and $'*'$ represents real, imaginary and conjugation operation respectively.

The injected voltage phasor $V_{SH}$ is treated as a free variable function of $\alpha$. The proposed holomorphic embedded system for (\ref{ST_Node}), (\ref{ST_Active_Cons}), and (\ref{ST_Modes}) are as follows:
\begin{multline}
V^{*}_{i}(\alpha^{*})\sum_{k=1}^{N} Y_{ik}V_{k}(\alpha) + V^{*}_{i}(\alpha^{*})\big \{ V_{i}(\alpha)+V_{SH}(\alpha)\big \}y_{SH}
 = \alpha \bigg [S^{*}_{i} - C^{*}_{i}\sum_{k=1}^{N} Y_{ik}C_{k} -C^{*}_{i}(C_{i}+C_{SH})y_{SH} \bigg ]
+ C^{*}_{i}\sum_{k=1}^{N} Y_{ik}C_{k} 
\\+ C^{*}_{i}(C_{i}+C_{SH})y_{SH}
\label{ST_Node_HE}
\end{multline}
\begin{equation}
\Re \big [ V_{SH}(\alpha) \big \{ V^{*}_{i}(\alpha^{*})+V^{*}_{SH}(\alpha^{*}) \big \} y^{*}_{SH} \big ] = - \alpha \Re \big [ C_{SH}(C^{*}_{i}+C^{*}_{SH})y^{*}_{SH} \big ] + \Re [C_{SH}(C^{*}_{i}+C^{*}_{SH})y^{*}_{SH}]
\label{ST_Active_Cons_HE}
\end{equation}
\begin{subequations}
\begin{equation}
 V_{i}(\alpha)V^{*}_{i}(\alpha^{*})  = C_{i}C^{*}_{i} + \alpha \big [ |V^{SP}_{i}|^2 - C_{i}C^{*}_{i} \big ] 
\end{equation} 
\begin{equation}
V_{SH}(\alpha)V^{*}_{SH}(\alpha^{*})  = C_{SH}C^{*}_{SH} + \alpha \big [ |V^{SP}_{SH}|^2 - C_{SH}C^{*}_{SH} \big ] 
\end{equation}
\begin{equation}
\Im \big [ V_{SH}(\alpha)\big \{ V^{*}_{i}(\alpha^{*})+V^{*}_{SH}(\alpha^{*}) \big \}y^{*}_{SH} \big ]  = \Im \big \{C_{SH}(C^{*}_{i}+C^{*}_{SH})y^{*}_{SH} \big \}+ \alpha \big [Q^{SP}_{SH} -\Im \big \{C_{SH}(C^{*}_{i}+C^{*}_{SH})y^{*}_{SH} \big \} \big ]  
\end{equation} 
\begin{equation}
\Im \big [ V_{i}(\alpha)\big \{ V^{*}_{i}(\alpha^{*})-V^{*}_{j}(\alpha^{*}) \big \}y^{*}_{ij} \big ]  =  \Im \big \{C_{i}(C^{*}_{i}-C^{*}_{j})y^{*}_{ij} \big \} +  \alpha \big [Q^{SP}_{ij} -\Im \big \{C_{i}(C^{*}_{i}-C^{*}_{j})y^{*}_{ij} \big \} \big ]
\end{equation} 
\begin{equation}
\Im \bigg [ \frac{\{ V_{i}(\alpha)+V_{SH}(\alpha)\}}{V_{SH}(\alpha)}y_{SH} \bigg ]  =  \Im \bigg \{ \frac{(C_{i}+C_{SH})y_{SH}}{C_{SH}} \bigg \}  + \alpha \bigg [b^{SP}_{eq(SH)} - \Im \bigg \{ \frac{(C_{i}+C_{SH})y_{SH}}{C_{SH}} \bigg \} \bigg ]
\end{equation} 
\label{ST_Modes_HE}
\end{subequations}

Equations (\ref{ST_Node_HE}), (\ref{ST_Active_Cons_HE}) and (\ref{ST_Modes_HE}) correspond to (\ref{ST_Node}), (\ref{ST_Active_Cons}), and (\ref{ST_Modes}) respectively and fulfil the requirement of embedding at the reference state $\alpha_{0}=0$ and the target state $\alpha_{1}=1$. The constants $C_{SH}\in \mathbb{C}\setminus \{0\}$ and $C_{k}\in \mathbb{C}\setminus \{0\}$ are adjustable and can be of any pre-specified values. The constant $C_{SH}$ is used to represent the initial value of injected voltage source $V_{SH}$, whereas constant $C_{k}$ represents the initial value of bus voltages. At $\alpha = 0$, the solution of (\ref{ST_Node_HE}) and (\ref{ST_Active_Cons_HE}) gives $V_{SH}[0] = C_{SH}$ and $V_{k}[0]=C_{k}$. A STATCOM can't inject active power into the system  but it can inject reactive power. Therefore, $n_{ST}$ STATCOMs can independently control $2n_{ST}-1$ parameters. Therefore, this configuration can control only one quantity of the power system independently.

\begin{figure*}[!t]
\begin{equation}
\resizebox{1\hsize}{!}{%
\rotatebox{0}{$ [A^{ST}] =\begin{bmatrix} \begin{array}{cccccccccc}   
1  	 &   0 	  & 0 &  0 & 0 & 0 & 0 & 0 & 0  & 0    \\
0  	 &   1    & 0 &  0 & 0 & 0 & 0 & 0 & 0  & 0   \\

\mu_{\mathcal{GF}} & \xi_{\mathcal{GF}} & \mu_{\mathcal{GG}} & \xi_{\mathcal{GG}} &  \mu_{\mathcal{GL}} & \xi_{\mathcal{GL}} &   \mu_{\mathcal{G}i} & \xi_{\mathcal{G}i}   & 0  & 0\\

 0 & 0 & C_{\mathcal{G}re} &  C_{\mathcal{G}im} & 0 & 0 &0 & 0 & 0  & 0  \\

\mu_{\mathcal{LF}} & \xi_{\mathcal{LF}} & \mu_{\mathcal{LG}} & \xi_{\mathcal{LG}} &  \mu_{\mathcal{LL}} & \xi_{\mathcal{LL}} &  \mu_{\mathcal{L}i}& \xi_{\mathcal{L}i}  & 0  & 0\\

 - \xi_{\mathcal{LF}} & \mu_{\mathcal{LF}} & -\xi_{\mathcal{LG}} & \mu_{\mathcal{LG}} &  \mu^{\bigstar}_{\mathcal{LL}} & \xi^{\bigstar}_{\mathcal{LL}} & - \xi_{\mathcal{L}i} & \mu_{\mathcal{L}i}  & 0 & 0\\

\mu_{i\mathcal{F}} & \xi_{i\mathcal{F}} & \mu_{i\mathcal{G}} & \xi_{i\mathcal{G}} &  \mu_{i\mathcal{L}}  & \xi_{i\mathcal{L}} &  \mu_{ii} + 2C_{ire}g_{SH}+C_{SHre}g_{SH} - C_{SHim}b_{SH} & \xi_{ii} + 2C_{iim}g_{SH}+C_{SHim}g_{SH} + C_{SHre}b_{SH}   & C_{ire}g_{SH} + C_{iim}b_{SH}  & C_{iim}g_{SH} - C_{ire}b_{SH} \\

- \xi_{i\mathcal{F}} & \mu_{i\mathcal{F}} & - \xi_{i\mathcal{G}} & \mu_{i\mathcal{G}} & - \xi_{i\mathcal{L}} & \mu_{i\mathcal{L}} &  \mu^{\bigstar}_{ii} + 2C_{ire}b_{SH}+C_{SHre}b_{SH} + C_{SHim}g_{SH} & \xi^{\bigstar}_{ii}  + 2C_{iim}b_{SH}+C_{SHim}b_{SH} - C_{SHre}g_{SH}   & C_{ire}b_{SH} - C_{iim}g_{SH}  & C_{iim}b_{SH} + C_{ire}g_{SH}\\

0 & 0 & 0 & 0 & 0 & 0 & C_{SHre}g_{SH} + C_{SHim}b_{SH} & C_{SHim}g_{SH} - C_{SHre}b_{SH} & C_{ire}g_{SH}-C_{iim}b_{SH} + 2C_{SHre}g_{SH}  & C_{iim}g_{SH}+C_{ire}b_{SH} + 2C_{SHim}g_{SH}\\

\multicolumn{10}{c}{ \cdots Select~row~vector~from~equation~(\ref{ST_Modes_Mat})~corresponding~to~desired~ control~mode \cdots} \\
\end{array} 
\end{bmatrix}$}}
\label{ST_Matrix}
\end{equation}
\begin{equation}
\resizebox{0.94\hsize}{!}{%
\rotatebox{0}{$\begin{bmatrix} \begin{array}{cccccccccc}   

 0  & 0 	 &   0 & 0 &  0 & 0 & C_{ire} & C_{iim} & 0 &  0  \\
 0  & 0      &   0 & 0 &  0 & 0 &  0 &  0 & C_{SHre}& C_{SHim} \\
 0  & 0      &   0 & 0 &  0 & 0 & C_{SEim}g_{SH}-C_{SHre}b_{SH} & -C_{SHre}g_{SH}-C_{SHim}b_{SH}  & -C_{iim}g_{SH}-C_{ire}b_{SH}-2C_{SHre}b_{SH} &  C_{ire}g_{SH}-C_{iim}b_{SH}-2C_{SHim}b_{SH}\\
 
 0  & 0      &   0 & 0 &  C_{ire}b_{i\mathcal{L}}-C_{iim}g_{i\mathcal{L}} & C_{iim}b_{i\mathcal{L}}+C_{ire}g_{i\mathcal{L}} & C_{\mathcal{L}re}b_{i\mathcal{L}}+C_{\mathcal{L}im}g_{i\mathcal{L}}-2C_{ire}b_{i\mathcal{L}}   & C_{\mathcal{L}im}b_{i\mathcal{L}}-C_{\mathcal{L}re}g_{i\mathcal{L}}-2C_{iim}b_{i\mathcal{L}}  &
 0 &  0  \\
 0  & 0      &   0 & 0 &  0 & 0 & \Im(\frac{y_{SH}}{C_{SH}})  &\Re(\frac{y_{SH}}{C_{SH}}) & -\Im(\frac{y_{SH}C_{i}}{C^{2}_{SH}}) &  -\Re(\frac{y_{SH}C_{i}}{C^{2}_{SH}})  \\
\end{array} 
\end{bmatrix} \begin{bmatrix}
       X^{ST}
        \end{bmatrix} =~\begin{bmatrix} \begin{array}{c}
        		   \Gamma_{M1}[n-1]  \\
        		   \Gamma_{M2}[n-1]  \\
        		   \Gamma_{M3}[n-1]  \\
                   \Gamma_{M4}[n-1]  \\
                   \Gamma_{M5}[n-1]  \\
 				\end{array} 
                \end{bmatrix}$}}
\label{ST_Modes_Mat} 
\end{equation}
\begin{equation}
\resizebox{0.9\hsize}{!}{%
\rotatebox{0}{$[X^{ST}] = \begin{bmatrix} \begin{array}{cccccccccc}
V_{\mathcal{F}re}[n] & V_{\mathcal{F}im}[n] & V_{\mathcal{G}re}[n] & V_{\mathcal{G}im}[n] & V_{\mathcal{L}re}[n] & V_{\mathcal{L}im}[n] & V_{ire}[n] & V_{iim}[n]  & V_{SHre}[n] & V_{SHim}[n]
\end{array}\end{bmatrix}^{'} $}} 
\label{ST_Vector}
\end{equation}
\begin{equation}
\resizebox{0.9\hsize}{!}{%
\rotatebox{0}{$[B^{ST}] = \begin{bmatrix} \begin{array}{cccccccccc}
\Re[\Gamma_{\mathcal{F}}[n-1]] & \Im[\Gamma_{\mathcal{F}}[n-1]] & \Gamma_{\mathcal{G}}[n-1] & \Gamma_{\mathcal{GV}}[n-1]  & \Re[\Gamma_{\mathcal{L}}[n-1]] & \Im[\Gamma_{\mathcal{L}}[n-1]] & \Re[\Gamma^{ST}_{i}[n-1]] & \Im[\Gamma^{ST}_{i}[n-1]]  & \Gamma^{ST}_{PBE}[n-1]         & \Gamma^{ST}_{Mi}[n-1]    \end{array}\end{bmatrix}^{'} $}} 
\label{ST_RHS}
\end{equation}
\hrulefill
\end{figure*}
The general recurrence relationships for $n\geq 1$ are obtained by comparing the coefficient of $\alpha^{n}$ and the system of linear equations for STATCOM $[A^{ST}]_{(\Upsilon\times\Upsilon)}[X^{ST}]_{(\Upsilon\times 1)}=[B^{ST}]_{(\Upsilon\times 1)}$ has been derived, where $\Upsilon=2(N+n_{ST})$. The derived coefficient matrix, unknown vector and known vector can be expressed as in (\ref{ST_Matrix}), (\ref{ST_Vector}) and (\ref{ST_RHS}) respectively (shown at the top of this page). The general recurrence formula for stated control modes are also formulated and shown in (\ref{ST_Modes_Mat}).

The some entries of known vector $[B^{ST}]$ can be expressed as follows:
\begin{multline}
\Gamma^{ST}_{i}[n-1] = \eta_{n1} \bigg [ S^{*}_{i} - C^{*}_{i} \sum^{N}_{k=1} Y_{ik}C_{k} - C^{*}_{i}(C_{i}+C_{SH})y_{SH} \bigg ] -  \sum^{n-1}_{d=1} \bigg \{ V^{*}_{i}[d]V_{i}[n-d]-V^{*}_{i}[d]V_{SH}[n-d]\bigg \}y_{SH}
\\- \sum^{N}_{k=1}  \sum^{n-1}_{d=1} Y_{ik} V^{*}_{i}[d]V_{k}[n-d]
\end{multline}
\begin{multline}
\Gamma^{ST}_{PBE}[n-1] = - \eta_{n1} \Re \big [ C_{SH}(C^{*}_{i}+C^{*}_{SH})y^{*}_{SH} \big ] - \Re \Bigg [  \sum^{n-1}_{d=1} \bigg \{ V_{SH}[d]V^{*}_{i}[n-d] + V_{SH}[d]V^{*}_{SH}[n-d]\bigg \}y^{*}_{SH} \Bigg ]
\end{multline}
\begin{equation}
\Gamma^{ST}_{M1}[n-1] = \frac{\eta_{n1}}{2} \big [(V^{SP}_{i})^{2}  - C_{i}C^{*}_{i} \big ] - \frac{1}{2}  \sum^{n-1}_{d=1} V_{i}[d]V^{*}_{i}[n-d] 
\label{BusV_Cons}
\end{equation}
The embedding of (\ref{ST_Mode1}) and (\ref{ST_Mode2}) are similar to each other, therefore, the general recurrence relationship for $\Gamma^{ST}_{M2}[n-1]$ can be obtained by changing the subscript $i$ to $SH$ in (\ref{BusV_Cons}).
\begin{equation}
\Gamma^{ST}_{M3}[n-1] = \eta_{n1}  \big [Q^{SP}_{SH} - \Im\big \{C_{SH}(C^{*}_{i}+C^{*}_{SH})y^{*}_{SH}\big\} \big ] - \Im \Bigg [  \sum^{n-1}_{d=1} \bigg \{ V_{SH}[d]V^{*}_{i}[n-d] + V_{SH}[d]V^{*}_{SH}[n-d]\bigg \}y^{*}_{SH} \Bigg ]
\end{equation}
\begin{equation}
\Gamma^{ST}_{M4}[n-1] = \eta_{n1}  \big [Q^{SP}_{ij} - \Im\big \{C_{i}(C^{*}_{i}-C^{*}_{j})y^{*}_{ij}\big\} \big ] - \Im \Bigg [  \sum^{n-1}_{d=1} \bigg \{ V_{i}[d]V^{*}_{i}[n-d] - V_{i}[d]V^{*}_{j}[n-d]\bigg \}y^{*}_{ij} \Bigg ]
\end{equation}
\begin{multline}
\Gamma^{ST}_{M5}[n-1] = \eta_{n1}  \Bigg [b^{SP}_{eq(SH)} - \Im\bigg \{\frac{(C_{i}+C_{SH})y_{SH}}{C_{SH}}\bigg\} \Bigg ] - \Im \Bigg [  \sum^{n-1}_{d=1} y_{SH} \bigg \{ -W_{SH}[d]V_{SH}[n-d]\frac{C_{i}}{C_{SH}} 
\\ +V_{i}[d]W_{SH}[n-d]\bigg \} \Bigg ]
\end{multline}

\subsection{UPFC Modeling}
\begin{figure}[b]
\centering
\includegraphics[scale=1]{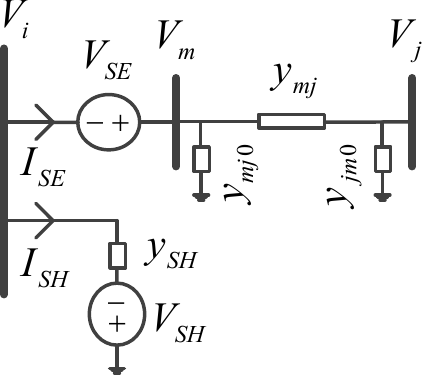}
\caption{Equivalent circuit of a UPFC}
\label{UPFC_Equivalent}
\end{figure}

A UPFC is a shunt-series connected FACTS device. Figure \ref{UPFC_Equivalent} shows the topological structure of a UPFC. It consists of two voltage source converters, two coupling transformers and a common capacitor. The shunt VSC is coupled to a local bus $i$ through a shunt connected transformer and the series VSC is coupled to a transmission line through a series connected transformer. The shunt converter can inject the reactive power into the system and it allows the exchange of the real power to the series converter through a common DC link to satisfy the device's physical constraint. The UPFC with $n_{UP}$ branches (\textit{i.e.} $2n_{UP}$ free variables) can independently control $2n_{UP}-1$ parameters because one variable is responsible for real power balance of the device. Therefore, the circuit configuration shown in Figure \ref{UPFC_Equivalent} can simultaneously control three quantities of the system independently. The real power can be exchanged amongst the shunt and series converters via a common DC link, but the sum of active power exchanged must be equal to zero.

Due to incorporation of the UPFC in the system, two new complex variable $V_{SH}$ and $I_{SE}$ have been introduced, which further add four unknown variables. Therefore, to find out the unique solution of unknown variables, four additional equations are required. To model the UPFC into load-flow algorithms, one equation from device's physical constraint and three equations from controlling modes of the device along with the modification of PBE at the local bus $i$ and receiving end bus $m$ are required. The mathematical formulation of the shunt and series converter controlling modes are similar to the controlling modes of STATCOM and SSSC \cite{Part1} respectively. The PBE at bus $i$, $m$ and real power power exchange constraints can be expressed as follows:
\begin{equation}
V^{*}_{i}\sum_{k=1}^{N} Y_{ik}V_{k} + V^{*}_{i}I_{SE} + V^{*}_{i}(V_{i}+V_{SH})y_{SH} = S^{*}_{i}
\label{UPFC_Node_S}
\end{equation}
\begin{equation}
V^{*}_{m}\sum_{k=1}^{N} Y_{mk}V_{k} - V^{*}_{m}I_{SE}= S^{*}_{m}
\label{UPFC_Node_R}
\end{equation}
\begin{equation}
\Re \big [ (V_{m}-V_{i})I^{*}_{SE} \big ] + \Re \big [ V_{SH} ( V^{*}_{i}+V^{*}_{SH})y^{*}_{SH} \big ] = 0
\label{UPFC_Active_Cons}
\end{equation} 
The following new embedded system is introduced for (\ref{UPFC_Node_S})-(\ref{UPFC_Active_Cons}):
\begin{multline}
V^{*}_{i}(\alpha^{*})\sum_{k=1}^{N} Y_{ik}V_{k}(\alpha) + V^{*}_{i}(\alpha^{*})\big \{ ( V_{i}(\alpha)+V_{SH}(\alpha))y_{SH}+ I_{SE}(\alpha)\big \}
 = C^{*}_{i}\sum_{k=1}^{N} Y_{ik}C_{k}+C^{*}_{i}\big \{  (C_{i}+C_{SH})y_{SH} + D_{SE} \big \} 
\\+ \alpha \Bigg [S^{*}_{i} - C^{*}_{i}\sum_{k=1}^{N} Y_{ik}C_{k}-C^{*}_{i} \big \{ (C_{i}+C_{SH})y_{SH}+D_{SE} \big \}\Bigg ]
\label{UPFC_Node_S_HE}
\end{multline}
\begin{equation}
V^{*}_{m}(\alpha^{*})\sum_{k=1}^{N} Y_{mk}V_{k}(\alpha) - V^{*}_{m}(\alpha^{*})I_{SE}(\alpha)= C^{*}_{m}\sum_{k=1}^{N} Y_{mk}C_{k}-C^{*}_{m}D_{SE}  + \alpha \Bigg [S^{*}_{m} - C^{*}_{m}\sum_{k=1}^{N} Y_{mk}C_{k}+C^{*}_{m}D_{SE} \Bigg ]
\label{UPFC_Node_R_HE}
\end{equation}
\begin{multline}
\Re \bigg [ \big \{V_{m}(\alpha)-V_{i}(\alpha)\big \}I^{*}_{SE}(\alpha^{*}) +  V_{SH}(\alpha) y^{*}_{SH} \big \{ V^{*}_{i}(\alpha^{*})+V^{*}_{SH}(\alpha^{*}) \big \}  \bigg ] = \Re \big [ \{C_{m}-C_{i}\}D^{*}_{SE} + C_{SH}(C^{*}_{i}+C^{*}_{SH})y^{*}_{SH}\big ]
\\ - \alpha \Re \big [ \{C_{m}-C_{i}\}D^{*}_{SE} + C_{SH}(C^{*}_{i}+C^{*}_{SH})y^{*}_{SH}\big ]
\label{UPFC_Active_Cons_HE}
\end{multline}
\begin{figure*}[b]
\hrulefill
\begin{equation}
\resizebox{1\hsize}{!}{%
\rotatebox{0}{$[A^{UP}] = \begin{bmatrix} \begin{array}{cccccccccccccc}   
1  	 &   0 	  & 0 &  0 & 0 & 0 & 0 & 0 & 0  & 0 & 0 & 0  & 0 & 0 \\
0  	 &   1    & 0 &  0 & 0 & 0 & 0 & 0 & 0  & 0 & 0 & 0  & 0 & 0 \\

\mu_{\mathcal{GF}} & \xi_{\mathcal{GF}} & \mu_{\mathcal{GG}} & \xi_{\mathcal{GG}} &  \mu_{\mathcal{GL}} & \xi_{\mathcal{GL}} &   \mu_{\mathcal{G}i} & \xi_{\mathcal{G}i} & \mu_{\mathcal{G}m} & \xi_{\mathcal{G}m}   & 0  & 0 & 0 & 0\\

 0 & 0 & C_{\mathcal{G}re} &  C_{\mathcal{G}im} & 0 & 0 &0 & 0 & 0  & 0  & 0 & 0 & 0 & 0 	  \\

\mu_{\mathcal{LF}} & \xi_{\mathcal{LF}} & \mu_{\mathcal{LG}} & \xi_{\mathcal{LG}} &  \mu_{\mathcal{LL}} & \xi_{\mathcal{LL}} &  \mu_{\mathcal{L}i}& \xi_{\mathcal{L}i} & \mu_{\mathcal{L}m} & \xi_{\mathcal{L}m} & 0  & 0 & 0  & 0\\

 - \xi_{\mathcal{LF}} & \mu_{\mathcal{LF}} & -\xi_{\mathcal{LG}} & \mu_{\mathcal{LG}} &  \mu^{\bigstar}_{\mathcal{LL}} & \xi^{\bigstar}_{\mathcal{LL}} & - \xi_{\mathcal{L}i} & \mu_{\mathcal{L}i} & - \xi_{\mathcal{L}m} & \mu_{\mathcal{L}m} &  0 & 0 & 0  & 0\\

\mu_{i\mathcal{F}} & \xi_{i\mathcal{F}} & \mu_{i\mathcal{G}} & \xi_{i\mathcal{G}} &  \mu_{i\mathcal{L}}  & \xi_{i\mathcal{L}} &  \mu_{ii} + D_{SE1re}+ D_{SEre}+ 2C_{ire}g_{SH}+C_{SHre}g_{SH} - C_{SHim}b_{SH}  & \xi_{ii}+ D_{SEim}+ 2C_{iim}g_{SH}+C_{SHim}g_{SH} + C_{SHre}b_{SH}  & \mu{im} & \xi_{im}  & C_{ire}g_{SH}+C_{iim}b_{SH}  & C_{iim}g_{SH}-C_{ire}b_{SH} & C_{ire}  & C_{iim}\\

- \xi_{i\mathcal{F}} & \mu_{i\mathcal{F}} & - \xi_{i\mathcal{G}} & \mu_{i\mathcal{G}} & - \xi_{i\mathcal{L}}  & \mu_{i\mathcal{L}} &  \mu^{\bigstar}_{ii} + D_{SEim}+ 2C_{ire}b_{SH}+C_{SHre}b_{SH} + C_{SHim}g_{SH}  & \xi^{\bigstar}_{ii}- D_{SEre}+  2C_{iim}b_{SH}+C_{SHim}b_{SH}-C_{SHre}g_{SH}  & -\xi_{im} & \mu_{im}   & C_{ire}b_{SH}-C_{iim}g_{SH}  & C_{ire}g_{SH}+C_{iim}b_{SH} & -C_{iim}  & C_{ire}\\

\mu_{m\mathcal{F}} & \xi_{m\mathcal{F}} & \mu_{m\mathcal{G}} & \xi_{m\mathcal{G}} &  \mu_{m\mathcal{L}}  & \xi_{m\mathcal{L}} &  \mu_{mi} &  \xi_{mi} & \mu_{mm} - D_{SEre} & \xi_{mm} - D_{SEim}  & 0 & 0 & - C_{mre}  & - C_{mim} \\

- \xi_{m\mathcal{F}} & \mu_{m\mathcal{F}} & - \xi_{m\mathcal{G}} & \mu_{m\mathcal{G}} & - \xi_{m\mathcal{L}}  & \mu_{m\mathcal{L}} &  -\xi_{mi} &  \mu_{mi} & \mu^{\bigstar}_{mm} - D_{SEim} & \xi^\bigstar_{mm} + D_{SEim} &     0 & 0 & C_{mim}  & - C_{mre} \\

0 & 0 & 0 & 0 & 0 & 0 & - D_{SE1re}- D_{SE2re} & - D_{SE1im}- D_{SE2im} & D_{SE1re}  & D_{SE2im}& C_{mre} - C_{ire} & C_{mim} - C_{iim}      & C_{tre} - C_{ire} & C_{tim} - C_{iim}\\

\multicolumn{14}{c}{ \cdots Select~3~different~row~vector~from~equation~(21)~of~[26]~ and~ (\ref{ST_Modes_Mat})~corresponding~to~desired~ control~mode \cdots} \\
\end{array} 
\end{bmatrix}$}}
\label{UPFC_Matrix}
\end{equation}
\begin{equation}
\resizebox{0.9\hsize}{!}{%
\rotatebox{0}{$[X^{UP}] = \begin{bmatrix} \begin{array}{cccccccccccccc}
V_{\mathcal{F}re}[n] & V_{\mathcal{F}im}[n] & V_{\mathcal{G}re}[n] & V_{\mathcal{G}im}[n] & V_{\mathcal{L}re}[n] & V_{\mathcal{L}im}[n] & V_{ire}[n] & V_{iim}[n] & V_{mre}[n] & V_{mim}[n] & V_{SHre}[n] & V_{SHim}[n] & I_{SEre}[n] & I_{SEim}[n]
\end{array}\end{bmatrix}^{'} $}} 
\label{UPFC_Vector}
\end{equation}
\begin{equation}
\resizebox{0.9\hsize}{!}{%
\rotatebox{0}{$[B^{UP}] = \begin{bmatrix} \begin{array}{cccccccccccccc}
\Re[\Gamma_{\mathcal{F}}[n-1]] & \Im[\Gamma_{\mathcal{F}}[n-1]] & \Gamma_{\mathcal{G}}[n-1] & \Gamma_{\mathcal{GV}}[n-1]  & \Re[\Gamma_{\mathcal{L}}[n-1]] & \Im[\Gamma_{\mathcal{L}}[n-1]] & \Re[\Gamma^{UP}_{i}[n-1]] & \Im[\Gamma^{UP}_{i}[n-1]]  & \Re[\Gamma^{UP}_{m}[n-1]] & \Im[\Gamma^{UP}_{m}[n-1]] & \Gamma^{UP}_{PBE}[n-1]         & \Gamma^{UP}_{Mi}[n-1]    & \Gamma^{UP}_{Mi}[n-1] & \Gamma^{UP}_{Mi}[n-1]  	  
\end{array}\end{bmatrix}^{'} $}} 
\label{UPFC_RHS}
\end{equation}
\end{figure*}

The similar procedure has been adopted to obtain the system of linear equations as discussed in Section \ref{STATCOM_modeling}. The system of linear equations for UPFC $[A^{UP}]_{(\Upsilon\times\Upsilon)}[X^{UP}]{(\Upsilon\times1)}=[B^{UP}]{(\Upsilon\times1)}$ is given in (\ref{UPFC_Matrix}), (\ref{UPFC_Vector}) and (\ref{UPFC_RHS}) respectively (shown at the bottom of this page), where $\Upsilon=2(N+2n_{UP})$. For purpose of numerical studies, the different controlling modes for the shunt and series converters have been adopted from (\ref{ST_Modes_Mat}) and \cite{Part1} respectively. Some entries of known vector $[B^{UP}]$ are directly taken from \cite{Part1} (\textit{e.g.} entries of slack, load and generator buses) and shown in Appendix A; and the remaining are as follows:

\begin{multline}
\Gamma^{UP}_{i}[n-1] = \eta_{n1} \bigg [ S^{*}_{i} - C^{*}_{i} \sum^{N}_{k=1} Y_{ik}C_{k} - C^{*}_{i}(C_{i}+C_{SH})y_{SH} - C^{*}_{i}D_{SE}\bigg ] -  \sum^{n-1}_{d=1} \bigg \{ V^{*}_{i}[d]V_{i}[n-d]+V^{*}_{i}[d]V_{SH}[n-d]\bigg \}y_{SH}
\\- \sum^{N}_{k=1}  \sum^{n-1}_{d=1} Y_{ik} V^{*}_{i}[d]V_{k}[n-d] - \sum^{n-1}_{d=1}V^{*}_{i}[d]I_{SE}[n-d]
\end{multline}
\begin{equation}
\Gamma^{UP}_{m}[n-1] = \eta_{n1} \bigg [ S^{*}_{m} - C^{*}_{m} \sum^{N}_{k=1} Y_{mk}C_{k} + C^{*}_{m}D_{SE} \bigg ]- \sum^{N}_{k=1}  \sum^{n-1}_{d=1} Y_{mk} V^{*}_{m}[d]V_{k}[n-d] +  \sum^{n-1}_{d=1} V^{*}_{m}[d]I_{SE}[n-d]  
\end{equation}
\begin{multline}
\Gamma^{UP}_{PBE}[n-1] = - \eta_{n1} \Re \big [ C_{SH}(C^{*}_{i}+C^{*}_{SH})y^{*}_{SH} + (C_{m}-C_{i})D^{*}_{SE} \big ] - \Re \Bigg [ \sum^{n-1}_{d=1} \bigg \{ V_{m}[d]I^{*}_{SE}[n-d]-V_{i}[d]I^{*}_{SE}[n-d] \bigg \} \Bigg ]
\\- \Re \Bigg [  \sum^{n-1}_{d=1} \bigg \{ V_{SH}[d]V^{*}_{i}[n-d] + V_{SH}[d]V^{*}_{SH}[n-d]\bigg \}y^{*}_{SH} \Bigg ] 
\end{multline}

\section{Results and Discussions} \label{results}
The proposed models of STATCOM and UPFC; and their multi-control capabilities have been tested on IEEE 118-bus test systems \cite{zimmerman2010matpower}. But for demonstration purpose, only selected numerical results on IEEE 118-bus test system have been presented in this paper. To validate the proposed models, investigation has been done by incorporating the different devices at different locations and by changing the target values of quantities. All the quantities are in \textit{p.u.} and the base power being is 100 MVA. The proposed FFHE based models have been tested and examined for all the control modes as discussed in Section \ref{PM}. The proposed FFHE and NR based models of STATCOM and UPFC were modelled in MATLAB environment and simulated on Intel(R) Core(TM) i3-4150 CPU 3.50 GHz processor with 4-GB RAM. For all the investigations, the mismatch tolerance of $10^{-8}$ is considered for maximum bus power mismatch. The values of unknown variables obtained after 3 iterations of NR method have been used as initial guesses for the FFHE load-flow method throughout this paper. The numerical results of IEEE 118-bus test system in the absence of FACTS devices are presented in Table \ref{Without_table}. The similar procedure as discussed in \cite{Part1} has been adopted to calculate the percentage reduction in mismatch and runtime.

\begin{table}[pos=b]
\centering
\caption{Numerical results of IEEE 118-bus test system without any FACTS device}
\begin{tabular}{ll}
\hline 
$V_{16}=0.9839\angle -17.80 \degree$ & $S_{16-17}=-0.1752-j0.0370$ \\
$V_{20}=0.9581\angle -17.81\degree$   & $S_{20-21}=-0.2869+j0.0496$ \\
$V_{75}=0.9675\angle -7.07\degree$   & $S_{75-74}=0.5236+j0.0604$ \\
$V_{114}=0.9607\angle -15.27\degree$  & $S_{114-115}=0.0134+j0.0088$ \\
\hline 
\end{tabular}
\label{Without_table}
\end{table}

\begin{table}[pos=b]
\centering
\caption{Numerical results when STATCOM is incoporated at bus no. 16 in the IEEE 118-bus test system}
\resizebox{0.99\textwidth}{!}{
\begin{tabular}{llccccc}
\hline 
& & Case 1 & Case 2 & Case 3  & Case 4  & Case 5  \\ 
\hline 
\multicolumn{2}{l}{Specified Parameters}  & $V^{SP}_{16}=1.1$ & $V^{SP}_{SH}=1$ & $Q^{SP}_{SH}=-0.5$ & $Q^{SP}_{16-17}=0$ &  $b^{SP}_{eq(SH)}=-0.8$ \\
\hline
\multirow{5}{*}{\rotatebox{90}{Power-flow }}\multirow{5}{*}{\rotatebox{90}{results}} & $V_{16}$ & $\textbf{1.1}\angle -20.28\degree$ &  $0.9977\angle -17.99\degree$ & $0.9519\angle-17.45\degree$ & $0.9867\angle-17.84\degree$ & $1.0324\angle-18.61\degree$  \\

& $V_{SH}$ & $1.1197\angle 158.68\degree$ & $\textbf{1}\angle 161.87\degree$ & $0.9466\angle 162.87\degree$ & $0.9872\angle 162.13\degree$ & $1.0407\angle 160.93\degree$ \\

& $S_{16-17}$ & $-0.2121+j0.6672$ & $-0.1769+j0.0578$ & $-0.1736-j0.1739$ & $-0.1754+\textbf{j0}$ & $-0.1843+j0.2506$ \\

& $S_{SH}$ & $0+j2.2210$  & $0+j0.2352$ & $0-\textbf{j0.5}$ & $0+j0.0470$ & $0+j0.8664$ \\

& $b_{eq(SH)}$ & $-1.7717$  & $-0.2352$ & $0.5580$ & $-0.0482$ &  $\textbf{-0.8}$ \\

\hline
\multicolumn{2}{c}{$\% \vartriangle E$}& $25.91$ & $24.24$ & $27.92$ & $28.95$ & $33.87$  \\
\hline
\multicolumn{2}{c}{$\%  T$}  & $11.63$ & $9.25$ & $8.10$ & $10.97$ & $ 9.23$ \\
\hline 
\end{tabular}} 
\label{ST_table1}
\end{table}

\begin{table}[pos=b]
\centering
\caption{Numerical results when STATCOM is incoporated at bus no. 114 in the IEEE 118-bus test system}
\resizebox{0.99\textwidth}{!}{
\begin{tabular}{llccccc}
\hline 
& & Case 1 & Case 2 & Case 3  & Case 4  & Case 5  \\ 
\hline 
\multicolumn{2}{l}{Specified Parameters}  & $V^{SP}_{114}=0.9$ & $V^{SP}_{SH}=1$ & $Q^{SP}_{SH}=-0.1$ & $Q^{SP}_{114-115}=0$ &  $b^{SP}_{eq(SH)}=0.2$ \\
\hline
\multirow{5}{*}{\rotatebox{90}{Power-flow }}\multirow{5}{*}{\rotatebox{90}{results}} & $V_{114}$ & $\textbf{0.9}\angle -15.03\degree$ &  $0.9927\angle -15.76\degree$ & $0.9569\angle-15.23\degree$ & $0.9598\angle-15.26\degree$ & $0.9539\angle-15.19\degree$  \\

& $V_{SH}$ & $0.8843\angle 165.96\degree$ & $\textbf{1}\angle 163.81\degree$ & $0.9559\angle 164.83\degree$ & $0.9595\angle 164.75\degree$ & $0.9520\angle 164.92\degree$ \\

& $S_{114-115}$ & $0.0095-j0.6308$ & $0.0083+j0.3854$ & $0.0135-j0.0319$ & $0.0134-\textbf{j0}$ & $0.0135-j0.0662$ \\

& $S_{SH}$ & $0-j1.3773$  & $0+j0.7374$ & $0-\textbf{j0.1}$ & $0-j0.0241$ & $0-j0.1813$ \\

& $b_{eq(SH)}$ & $1.7613$  & $-0.7374$ & $0.1094$ & $0.0262$ &  $\textbf{0.2}$ \\

\hline
\multicolumn{2}{c}{$\% \vartriangle E$}& $22.35$ & $23.16$ & $28.48$ & $24.23$ & $29.18$  \\
\hline
\multicolumn{2}{c}{$\%  T$}  & $12.43$ & $9.47$ & $7.69$ & $9.78$ & $ 10.89$ \\
\hline 
\end{tabular}} 
\label{ST_table2}
\end{table}

\newpage
To validate the proposed embedding for STATCOM and its controlling modes, several test have been carried out on IEEE 118-bus test system. For demonstration purpose, only two location results have been presented in Table \ref{ST_table1}. For examining cases 1-5, STATCOM is assumed to be connected at bus 16. The specified reference values; and calculated values of STATCOM's variables and selected system variables are given in row 2 and 3 respectively. In case 1, the voltage magnitude of STATCOM connected bus was chosen as a control variable. In this case, STATCOM is able to set the voltage of bus 16 to the specified reference 1.1 \textit{p.u.} by providing the reactive power support to that bus. The examination of case 2 from this table proves that the proposed formulation also enables the model of STATCOM to control its injected voltage to a specified reference 1 \textit{p.u.} Case 3 demonstrates the reactive power absorption capability of the device and it can be observed that the absorbed reactive power is equal to the desired value. As per literature, STATCOM also has the capability to control reactive power-flow through the line, case 4 verifies the same. In this case, STATCOM controls the reactive power-flow of line 16-17 from $-$0.0370 \textit{p.u.} to target value 0 \textit{p.u.} So, by forcing the reactive power-flow through the line to zero, the active power transferring capability of the line can be increased. This particular mode may be attractive in context of the electricity market environment. The proposed formulation also enables the model to operate like a controllable admittance and it can be verified from case 5 as shown in Table \ref{ST_table1}.

Now, a STATCOM is installed at bus 114 and the desired values of control parameters, calculated values of system variables, STATCOM's parameters for all cases are presented in Table \ref{ST_table2}. From this table, it can be observed that the all the desired values (shown in bold letters) have been achieved. From Tables \ref{ST_table1} and \ref{ST_table2}, it can also be verified that the NR based model of STATCOM takes more runtime as compared to NR-assisted FFHE based model of STATCOM and the proposed model also reduces the error at faster rate. In the literature only STATCOM model based on standard HELM has been developed in \cite{singh2019statcom}, therefore, the performance of NR-assisted FFHE and basic HELM based model of STATCOM is also investigated. From the various test cases, it has been observed that the FFHE based model of STATCOM takes 80-90\% less time and requires 40-70\% less terms to converge as compared to basic HELM based model of STATCOM. 

\begin{table}[pos=b]
\centering
\caption{Numerical results when UPFC in the IEEE 118-bus test system (location: $75-74$) }
\begin{tabular}{llcc}
\hline 
&  & Case 1 & Case 2  \\ 
\hline 
\multicolumn{2}{l}{\multirow{3}{1cm}{Specified Parameters}}  & $V^{SP}_{75}=1$ & $V^{SP}_{SH}=1$  \\
                                 &         & $P^{SP}_{75-74}=0.75$ & $P^{SP}_{75-74}=0.2$  \\
                                 &         & $Q^{SP}_{75-74}=0 $ & $Q^{SP}_{75-74}=0.1$  \\
\hline                                          
\multirow{9}{*}{\rotatebox{90}{Power-flow results}} & $V_{75}$ & $\textbf{1}\angle -7.99\degree$ &  $0.9937\angle -6.80\degree$  \\

& $V_{SH}$ & $1.0087\angle 171.52\degree$ & $\textbf{1}\angle 172.83\degree$ \\
& $S_{SH}$ & $0.0129+j0.8636$  & $-0.002+j0.6315$  \\
& $b_{eq(SH)}$ & $-0.8488$  & $-0.6315$  \\
& $S_{75-74}$ & $\textbf{0.75+j0}$ & $\textbf{0.2+j0.1}$  \\
& $V_{SE}$ & $0.0725\angle 95.78\degree$ & $0.0855\angle -117.30\degree$ \\
& $I_{SE}$ & $0.75\angle -7.99\degree$ & $0.2250\angle -33.37\degree$ \\
& $S_{SE}$ & $-0.0129+j0.0528$  & $0.002-j0.0191$  \\
& $X_{eq(SE)}$ & $0.0938$  & $-0.3778$  \\
\hline
\multicolumn{2}{c}{$\% \vartriangle E$}  & $26.58$ & $23.37$  \\
\hline
\multicolumn{2}{c}{$\%  T$}             & $5.19$ & $6.45$  \\
\hline 
\end{tabular}
\label{UPFC_table1}
\end{table}

Further, the proposed embedded system for UPFC has been also tested for two cases. Firstly, it is assumed that the UPFC is connected between the buses 75 and 74 at the location of bus 75. In case 1, the shunt control of UPFC is to control the voltage at bus 75 and the series controls of UPFC are to control active and reactive power flowing through the transmission line 75-74. Similarly, in case 2, the shunt control of UPFC is to control injected voltage and series control responsibilities of UPFC are the same except the desired values. The rows 2, 3, 4, 5 of Table \ref{UPFC_table1} presents the specified parameters, calculated values of UPFC's and system variables, percentage reduction in error and runtime respectively. The active power exchange between the shunt and series converters is also presented in this table. In Table \ref{UPFC_table1}, numerically it is shown that the all target values (shown in bold letters) have been met for both cases. The decrement in error ranges from 15\% to 27\% and reduction in runtime ranges from 3\% to 7\%.

Table \ref{UPFC_table2} presents the results for two cases when the UPFC is assumed to be connected between buses 20 and 21. The UPFC parameters \textit{i.e.} magnitudes and phase angles of the UPFC’s injected voltages $V_{SE}$, $V_{SH}$ are also given in this table, which enforce the system operating point to set as per specified reference quantities. The calculated values of specified references are shown in bold letters and it can be verified that all the desired values have been achieved. From Tables \ref{UPFC_table2}, it can be observed that the error reduction is higher and runtime is lesser for FFHE based model of UPFC. Note that in all the cases, convergence has been achieved; and the proposed models exhibited very good convergence and also converge at faster rate as compared to models based on standard NR method.

\begin{table}[pos=t]
\centering
\caption{Numerical results when UPFC in the IEEE 118-bus test system (location: $20-21$) }
\begin{tabular}{llcc}
\hline 
&  & Case 1 & Case 2  \\ 
\hline 
\multicolumn{2}{l}{\multirow{3}{1cm}{Specified Parameters}}  & $V^{SP}_{20}=1$ & $V^{SP}_{SH}=1$  \\
                                 &         & $P^{SP}_{20-21}=0.6$ & $P^{SP}_{20-21}=-0.4$  \\
                                 &         & $Q^{SP}_{20-21}=0 $ & $Q^{SP}_{20-21}=0$  \\
\hline                                          
\multirow{9}{*}{\rotatebox{90}{Power-flow results}} & $V_{20}$ & $\textbf{1}\angle -28.95\degree$ &  $0.9981\angle -17.48\degree$  \\

& $V_{SH}$ & $1.0051\angle 150.77\degree$ & $\textbf{1}\angle 162.38\degree$ \\
& $S_{SH}$ & $0.0093+j0.5087$  & $-0.0290+j0.2182$  \\
& $b_{eq(SH)}$ & $-0.5035$  & $-0.2182$  \\
& $S_{20-21}$ & $\textbf{0.6+j0}$ & $\textbf{$-$0.4+j0}$  \\
& $V_{SE}$ & $0.5545\angle 62.65\degree$ & $0.0866\angle -164.27\degree$ \\
& $I_{SE}$ & $0.6\angle -28.95\degree$ & $0.4008\angle 162.53\degree$ \\
& $S_{SE}$ & $-0.0093+j0.3326$  & $0.0290+j0.0190$  \\
& $X_{eq(SE)}$ & $0.9238$  & $0.1183$  \\
\hline
\multicolumn{2}{c}{$\% \vartriangle E$}  & $24.31$ & $25.20$  \\
\hline
\multicolumn{2}{c}{$\%  T$}             & $7.54$ & $6.01$  \\
\hline 
\end{tabular}
\label{UPFC_table2}
\end{table}
\begin{table}[pos=t]
\centering
\caption{Power-flow results when device limit constraints are imposed  }
\begin{tabular}{cccl}
\hline 
Device  & Specified Parameter (\textit{p.u.}) & Limit Violated & Power-flow results \\
\hline
\multirow{7}{*}{STATCOM}& \multirow{7}{*}{$V^{SP}_{16}=1.1$} & \multirow{7}{*}{Yes} & $V_{16}=\textcolor{blue}{1.0830}\angle-19.81\degree$\\
                                         &                   &     & $V_{SH}=\textbf{1.1}\angle159.28\degree$\\
                                         &                   &     & $S_{16-17}=-0.2038+j0.5580$\\
                                         &                   &     & $S_{SH}=0+j1.8806$\\
                                         &                   &     & $b_{eq(SH)}=-1.5542$\\
                                         &                   &     & $\% \vartriangle E=23.85$\\
                                         &                   &     & $\%  T=6.38$\\

\hline 
\multirow{11}{*}{UPFC}   &               & \multirow{11}{*}{Yes} & $V_{20}=\textbf{1}\angle-23.77\degree$\\
                                         &                   &     & $V_{SE}=\textbf{0.3}\angle72.46\degree$\\
                                         &                   &     & $I_{SE}=0.1813\angle-23.77\degree$\\
                                         &                   &     & $S_{20-121}=\textcolor{blue}{0.1813}+j\textbf{0}$\\
           &    $V^{SP}_{20}=1$                              &     & $S_{SE}=-0.0059+j0.0541$\\
          &$P^{SP}_{20-21}=0.6$                             &     & $X_{eq(SE)}=1.6446$\\
          &    $Q^{SP}_{20-21}=0$                            &     & $V_{SH}=1.0038\angle156.63$\\
                                         &                   &     & $S_{SH}=0.0059+j0.3724$\\
                                         &                   &     & $b_{eq(SH)}=-0.3696$\\
                                         &                   &     & $\% \vartriangle E=17.37$\\
                                         &                   &     & $\%  T=3.94$\\
 \hline 
\end{tabular}
\label{Limits_table}
\end{table}

The introduced embedded system has also been investigated with device limit constraints. To demonstrate the handling of constraints, one case from Table \ref{ST_table1} and another case from Table \ref{UPFC_table2} were chosen. The maximum value of injected voltage by SSSC and STATCOM were selected as 0.3 \textit{p.u.} and 1.1 \textit{p.u.} respectively. From Table \ref{ST_table1}, it can be observed that when the target voltage of bus 16 is chosen as 1.1 \textit{p.u.} and no limits were set on the injected voltage, the STATCOM is able to control the voltage magnitude of bus 16 to the specified reference. But when the limits are imposed, STATCOM is not able to control the same because higher $V_{SH}$ is required to maintain the same. Therefore, whenever the limit constraints are violated, current mode is relaxed; and STATCOM will acts as a constant voltage source (\textit{i.e.} injected voltage magnitude mode is activated) and reference value is set equal to the value of violated limit. From Table \ref{Limits_table}, it can be observed that the voltage magnitude of bus 16 is not equal to
the reference value but it is equal to the 1.0830 \textit{p.u.}, while the injected voltage is maintained to 1.1 \textit{p.u.} Table \ref{Limits_table} also presents the results when the limit constraints for UPFC are imposed and similar results are achieved. Therefore, it can be concluded that the proposed FFHE based models of STATCOM and UPFC are also suitable when limit constraints of the devices are imposed.

The NR and FFHE based models have been tested on same platform for comparison, although the derived inferences may not be general due to various differences in simulation structure, programming skills \textit{etc.} It is observed that the percentage reduction in error and runtime are different for each case. Because there is no particular choice of the initial guess for the proposed models to ensure fast convergence in all cases. The selection of promising initial guess still remains an issue. Briefly, the runtime and rate of convergence of the proposed models will vary for different initial guess. Although, the performance of proposed FFHE based FACTS devices model are sensitive to the initial guess, but the proposed model perform better than basic HELM based models. Moreover, for most of cases, FFHE based models outperforms NR based models in context of error reduction. A common conclusion for all the stated devices and their controlling modes is that the proposed embedding for the devices are reliable and have good convergence characteristics. Further, the proposed formulation reduces the mismatch error at faster rate and therefore, converges slightly faster than the NR method based FACTS devices models. Therefore, the proposed FFHE based models offers a good alternative to the NR based models when promising initial guesses are available.  

\section{Conclusion} \label{conclusions}
The purpose of this research work is to develop flexible HELM models of STATCOM, and UPFC. To do the same, the power balance equations and devices' controlling modes are embedded with a complex variable $\alpha$ in such a manner that the embedded equations satisfy the requirement of embedding. Afterwards, the unknown variables were represented by power series and recursive formulas have been obtained for $n\geq1$. Lastly, at $\alpha=1$ numerical values of unknown variables are calculated using determinant method. The proposed embedded system of stated devices provides flexibility because any state can serve as an initial guess.  

From the numerical results, it is observed that the results of FFHE based models are similar to results obtained by NR based model of devices in perspective of the final calculated values of devices parameters, system variables and operational bounds. The comparison of runtime for FFHE based models with NR based models showed that later one took more time. Although the performance of FFHE based models is sensitive to the initial guess but the NR assisted FFHE based models outperform the NR based models in error reduction. The proposed model of STATCOM takes very less execution time as compared to the basic HELM model of STATCOM. The numerous results showed that the FFHE based models represents a step forward compared to the NR based models. So, the proposed models may be very useful at the planning stage of the power system. Therefore, the proposed FFHE based models offers a good alternative to the NR based models when promising initial guesses are available.

In this paper, no sophisticated technique is used during implementation of FFHE based models, therefore, possibility of reducing runtime further is still large via parallel computing, code optimization \textit{etc.}

\appendix
\section{Entries of Coefficient matrix and Known Vector}
The entries of coefficient matrix $[A^{ST}]$ and $[A^{UP}]$ are as follows:
\begin{equation}
\mu_{ik} = G_{ik}C_{ire} + B_{ik}C_{iim}~~;~ k\neq i 
\end{equation}
\begin{equation}
\xi_{ik} = G_{ik}C_{iim} - B_{ik}C_{ire}~~;~ k\neq i
\end{equation}
\begin{equation}
\mu_{ii} = G_{ii}C_{ire} + B_{ii}C_{iim} + \sum^{N}_{k=1} \big ( G_{ik}C_{kre} - B_{ik} C_{kim} \big ) 
\end{equation}
\begin{equation}
\xi_{ii} = G_{ii}C_{iim} - B_{ii}C_{ire} + \sum^{N}_{k=1} \big ( G_{ik}C_{kim} + B_{ik} C_{kre} \big )
\end{equation}
\begin{equation}
\mu^{\bigstar}_{ii} = - G_{ii}C_{iim} + B_{ii}C_{ire} + \sum^{N}_{k=1} \big ( G_{ik}C_{kim} + B_{ik} C_{kre} \big ) 
\end{equation}
\begin{equation}
\xi^{\bigstar}_{ii} = G_{ii}C_{ire} + B_{ii}C_{iim} + \sum^{N}_{k=1} \big ( B_{ik}C_{kim} - G_{ik} C_{kre} \big ) 
\end{equation}
The entries of known vector $[B^{ST}]$ and $[B^{UP}]$ are as follows:
\begin{equation}
\Gamma_{\mathcal{F}} = \eta_{n1}(V^{SP}_{\mathcal{F}}-C_{\mathcal{F}})
\end{equation}
\begin{equation}
\Gamma_{\mathcal{G}} = \frac{\eta_{n1}}{2} \Bigg [ 2P_{\mathcal{G}} - \Re \bigg \{C_{\mathcal{G}}\sum^{N}_{k=1}Y^{*}_{\mathcal{G}k}C^{*}_k \bigg \} \Bigg ] - \Re \Bigg [ \sum^{N}_{k=1}  \sum^{n-1}_{d=1} Y^{*}_{\mathcal{G}k} V^{*}_{\mathcal{G}}[d]V_{k}[n-d] \Bigg ]
\end{equation}
\begin{equation}
\Gamma_{\mathcal{GV}} = \frac{\eta_{n1}}{2} \big [(V^{SP}_{\mathcal{G}})^{2}  - C_{\mathcal{G}}C^{*}_{\mathcal{G}} \big ] - \frac{1}{2}  \sum^{n-1}_{d=1} V_{\mathcal{G}}[d]V^{*}_{\mathcal{G}}[n-d] 
\label{PV_Cons}
\end{equation}
\begin{equation}
\Gamma_{\mathcal{L}} = \eta_{n1} \bigg [ S^{*}_{\mathcal{L}} - C^{*}_{\mathcal{L}} \sum^{N}_{k=1} Y_{\mathcal{L}k}C_{k} \bigg ]
 - \sum^{N}_{k=1}  \sum^{n-1}_{d=1} Y_{\mathcal{L}k} V^{*}_{\mathcal{L}}[d]V_{k}[n-d] 
\end{equation}

\begin{equation}
\Gamma^{SC}_{M1} = \eta_{n1}  \big [P^{SP}_{im} - \Re(C_{i}D^{*}_{SE}) \big ] - \Re \Bigg [  \sum^{n-1}_{d=1} V_{i}[d]I^{*}_{SE}[n-d]  \Bigg ]
\end{equation}
\begin{equation}
\Gamma^{SC}_{M2} = \eta_{n1}  \big [Q^{SP}_{im} - \Im(C_{i}D^{*}_{SE}) \big ] - \Im \Bigg [  \sum^{n-1}_{d=1} V_{i}[d]I^{*}_{SE}[n-d]  \Bigg ]
\end{equation}
As, $F_{SE}(\alpha)$ and $|I_{SE}(\alpha)|$ are the inverse and magnitude of $I_{SE}(\alpha)$, so, the general recurrence relationship between these can be expressed as given in (\ref{inverse}) and (\ref{mag}).
\begin{equation}
F_{SE}[n] = \frac{-1}{D_{SE}}\sum^{n-1}_{d=0}F_{SE}[d]I_{SE}[n-d]
\label{inverse}
\end{equation}
\begin{equation}
\big|I_{SE}[n]\big| = \frac{1}{2\big|D_{SE}\big|}\bigg\{ \sum^{n}_{d=0}I_{SE}[d]I^{*}_{SE}[n-d]  - \sum^{n-1}_{d=1}\big|I_{SE}[d]\big|\big|I_{SE}[n-d]\big|\bigg\}
\label{mag}
\end{equation}

\bibliographystyle{elsarticle-num}


\bibliography{FFHEREF}


%
%

\end{document}